\documentclass[twocolumn,showpacs,superscriptaddress]{revtex4-1} 

\usepackage{amssymb}
\usepackage{subfigure}
\usepackage{amsmath}

\usepackage{graphicx}
\usepackage{epstopdf}
\usepackage{color}
\usepackage{ulem}

\newcommand{\scl}{0.10} 

\begin{document}

\title{Exceptional Points in Two Dissimilar Coupled Diode Lasers}

\date{\today}

\author{Yannis Kominis}
\affiliation{School of Applied Mathematical and Physical Science, National Technical University of Athens, Athens, Greece	}

\author{Kent D. Choquette}
\affiliation{Department of Electrical and Computer Engineering, University of Illinois, Urbana, Illinois, USA}

\author{Anastassios Bountis}
\affiliation{Department of Mathematics, School of Science and Technology, Nazarbayev University, Astana, Republic of Kazakhstan}

\author{Vassilios Kovanis}
\affiliation{Department of Physics, School of Science and Technology, Nazarbayev University, Astana, Republic of Kazakhstan}

\begin{abstract}
We show the abundance of Exceptional Points in the generic asymmetric configuration of two coupled diode lasers, under nonzero optical detuning and differential pumping. We pinpoint the location of these points with respect to the stability domains and the Hopf bifurcation points, in the solution space as well as in the space of experimentally controlled parameters.
\end{abstract}

\maketitle

\section{Introduction}
Coupled semiconductor laser arrays have been a subject of intense theoretical and experimental research for more than four decades, due to their capabilities for numerous applications as high-power laser sources, tunable photonic oscillators, controllable optical beam shaping and steering elements and ultrasensitive sensors \cite{Choquette_13, Choquette_15, Choquette_17, Choquette_18}. The coherent optical coupling of semiconductor lasers has been shown to lead to a large variety of interesting nonlinear dynamical features including the existence of phase-locked states, instabilities and bifurcations, localized synchronization and existence of complex states \cite{Wang_88, Winful&Wang_88, Winful_90, Otsuka_90, Winful_92, Kovanis_97, Hizanidis_17}. The underlying model of such structures is a system of coupled rate equations governing the time evolution of the electric fields amplitudes and phases as well as the carrier dynamics, with the carrier-induced nonlinearity playing a crucial role due to a non-zero linewidth enhancement factor describing an amplitude-phase coupling mechanism \cite{Erneux_book}.\

Although the dynamics of such structures under symmetric pumping are well known for more than two decades \cite{Wang_88, Winful&Wang_88, Winful_90, Otsuka_90, Winful_92}, it is quite recently that it has been shown that the utilization of asymmetric (optical or electrical) pumping schemes can control the characteristics of the phase-locked states such as their electric field ratio and phase difference \cite{Kominis_17a}. Moreover it has been shown that the eigenvalues of these states include Hopf bifurcations and Exceptional Points that control the spectral lineshape of the emitted light \cite{Kominis_17b}. The existence of Exceptional Points is also crucial for capabilities for ultrasensitive sensors \cite{Yang_16, Yang_17, Khajavikhan_17a, Khajavikhan_17b}.

For a long time the existence of Exceptional Points has been attributed to Parity-Time (PT) symmetry properties of the system, when simplified coupled mode equation models, neglecting the significant effects of carrier-induced nonlinearities, have been considered \cite{Christodoulides_10, Christodoulides_2016, Konotop_16}, with PT-symmetry neccesitating a zero frequency detuning between the coupled lasers. In a recent paper \cite{Choquette_18} it has been shown that unbroken PT-symmetry can be achieved even for non-zero frequency detuning by judiciously choosing unequal pumping rates, when the model of coupled rate equations is considered. In such case an effective PT-symmetry can be achieved and the Exceptional Points can be found at the boundary between broken and unbroken PT-symmetry. However, operating the coupled laser system at Exceptional Points can be challenging due to experimental difficulties related to the simultaneous control of the different pumping rates and the frequency detuning. The latter are related to the fact that the magnitude of the injection currents not only change the pump parameters but also varies the cavity resonance frequency through ohmic heating and refractive index temperature dependence \cite{Choquette_13, Choquette_18}.\

In this manuscript we show that Exceptional Points exist under a larger set of configurations, including substantial differential pumping and non-zero induced optical frequency detuning. This results in phase-locked states that are neither PT-symmetric nor effectively PT-symmetric, thus extending significantly the range of options for operating the system at an Exceptional Point. Moreover, we demonstrate the exact location of the Exceptional Points in the solution space and in the parameter space (evanescent coupling, frequency detuning and differential pumping), facilitating the appropriate parameter selection even when specific experimental constraints have to be considered. \

This paper is organized as follows: Section II presents the working model and the existence of asymmetric phase-locked states. In Section III, the eigenvalue spectrum of the phase-locked states is studied and the characteristic Hopf bifurcation and Exceptional points are located in the solution and in the parameter space. In Section IV, we present our concluding remarks and future work is proposed.

\section{Working model and asymmetric phase-locked states}
The dynamics of an array of two evanescently coupled semiconductor lasers is governed by the following coupled single-mode rate equations for the amplitude of the normalized electric fields $\mathcal{X}_i$, the phase difference $\theta$ between the fields and the normalized excess carrier density $Z_i$ of each semiconductor laser:
\begin{eqnarray}
 \dot{X}_1&=&X_1Z_1-\Lambda X_2\sin\theta \nonumber \\
 \dot{X}_2&=&X_2Z_2+\Lambda X_1\sin\theta \nonumber \\
 \dot{\theta}&=&\Delta -\alpha(Z_2-Z_1)+\Lambda\left(X_1/X_2-X_2/X_1\right)\cos\theta   \label{pair} \\
T\dot{Z}_1&=&P_1- Z_1-(1+2 Z_1)X_1^2 \nonumber \\
T\dot{Z}_2&=&P_2- Z_2-(1+2 Z_2)X_2^2 \nonumber
\end{eqnarray}
Here $\alpha$ is the linewidth enhancement factor, $\Lambda$ is the normalized coupling constant, $P_i$ is the normalized excess pumping rate, $\Delta=\omega_2-\omega_1$ is the cavity detuning, $T$ is the ratio of carrier to photon lifetimes, and the dot denotes derivative with respect to time which is normalized to the photon lifetime $\tau_p$ \cite{Winful&Wang_88, Choquette_13}. \

The phase-locked states of the system (\ref{pair}), are given by setting the time derivatives of the system equal to zero and their stability is determined by the eigenvalues of the Jacobian of the linearized system. For the case of zero detuning ($\Delta =0$) and symmetric pumping ($P_1=P_2=P_0$), two phase-locked states are known analytically: $X_1=X_2=\sqrt{P_0}$, $Z_1=Z_2=0$ and $\theta=0,\pi$. The in-phase state ($\theta=0$) is stable for $\Lambda>\alpha P_0 /(1+2P_0)$ whereas the out-of-phase state ($\theta=\pi$) is stable for $\Lambda<(1+2P_0)/2\alpha T$ \cite{Winful&Wang_88}. Although, phase-locked states, under general conditions  on nonzero detuning and unequal pumping, cannot be found analytical, in a recent work \cite{Kominis_17a} it has been shown that it is possilbe to derive  analytical expressions for the parameters of the system in terms of characteristic quantities of the phase-locked states, such as the field amplitude ratio $\rho \equiv X_2/X_1$ and phase difference $(\theta)$. This \textit{reverse approach} yields simple expressions for the pumping rates $P_i$ and detuning $\Delta$:
\begin{eqnarray}
 \Delta&=&-\alpha \Lambda\sin\theta\left(1/\rho+\rho\right)-\Lambda\cos\theta\left(\rho^{-1}-\rho\right)  \label{D_eq}\\
 P_1&=&X_0^2+(1+2X_0^2)\Lambda\rho\sin\theta  \nonumber\\
 P_2&=&\rho^2X_0^2-(1+2\rho^2 X_0^2) \Lambda\rho^{-1}\sin\theta  \label{P_eq}
\end{eqnarray}
as well as for the values of $Z_i$ corresponding to the phase-locked state:
\begin{eqnarray}
Z_1&=&\Lambda \rho \sin\theta \nonumber \\
Z_2&=&-\Lambda\rho^{-1}\sin\theta \label{Z_eq}
\end{eqnarray} 
where is the reference electric field amplitude $X_0=X_1$. The above equations clearly show that $P_i$ and $\Delta$ can be chosen to obtain a phase-locked state of arbitrary amplitude asymmetry $\rho$, phase difference $\theta$ and reference amplitude $X_0$. In the case of zero detuning ($\Delta=0$) between the coupled lasers the phase difference is restricted as 
\begin{equation}
\theta=s\pi + \tan^{-1}\left[\frac{1}{\alpha} \frac{\rho^2-1}{\rho^2+1}\right], \hspace{2em} s=0,1 \label{theta_rho} \\ 
\end{equation}
whereas in the case of symmetrically pumped lasers ($P_1=P_2=P_0$) the reference amplitude $X_0$ is fixed as
\begin{equation}
X_0^2 = \frac{ \Lambda \sin\theta (\rho^2+1)}{\rho\left[(\rho^2-1)-4 \Lambda \rho \sin\theta\right]} \label{X0_rho}
\end{equation}
and the common pumping rate is $P_0=X_0^2+(1+2X_0^2) \Lambda \rho \sin\theta$. It is worth emphasizing that even under a completely symmetric configuration ($\Delta=0$, $P_1=P_2$) asymmetry emerges in the form of a stable asymmetric phase-locked state which apparently has escaped the attention of  researchers for nearly 30 years \cite{Winful&Wang_88, Kominis_17a}. 

It is worth comparing the asymmetric phase-locked states with PT-symmetric or effectively PT-symmetric states as recently found in \cite{Choquette_18}. As seen from the form of Eqs. (\ref{pair}) and (\ref{D_eq})-(\ref{Z_eq}) we can define an effective frequency detuning $\Delta \omega$, the gain contrast $\Delta \gamma$ and the net gain $\delta$, corresponding to a phase-locked state as follows:
\begin{equation}
\Delta \omega \equiv \alpha (Z_2-Z_1)-\Delta =\Lambda \cos \theta \left( \rho -\rho^{-1} \right) 
\end{equation}
\begin{equation}
\Delta \gamma \equiv Z _2-Z_1=-\Lambda \sin \theta \left( \rho +\rho^{-1}  \right) 
\end{equation}
\begin{equation}
\delta \equiv Z_1+Z_2=\Lambda \sin \theta \left( \rho -\rho^{-1} \right) 
\end{equation}
The effective PT-symmetry corresponds to $\Delta \omega=0$, which can happen only for $\rho=1$, that is for symmetric phase-locked states. In such case we also have $\delta=0$, that is zero net gain. In all other cases where $\rho \neq 1$ both $\Delta \omega$ and  $\delta$ are non-zero, and the system is rendered non PT-symmetric in any sense. 

\section{Hopf Bifurcation and Exceptional Points of Asymmetric Phase-Locked States}
The eigenvalue spectrum of the asymmetric phase-locked states determines their stability and their spectral line shape. The Hopf Bifurcation points and the Exceptional Points play a crucial role, with the former corresponding to intensity peaks and the latter corresponding to side band merging \cite{Kominis_17b} . The eigenvalues of the phase-locked modes are calculated from the Jacobian of the dynamical system (\ref{pair}):

\begin{equation}
J=\left[
  \begin{array}{cc}
  A_{3\times3} & B_{3\times2} \\
  C_{2\times3} & D_{2\times2}
  \end{array}
  \right] 
\end{equation}
where
\begin{equation}
A_{3\times3}=\left[
  \begin{array}{ccc}
  \Lambda \rho \sin\theta & -\Lambda \sin\theta & -\Lambda \rho X_0 \cos \theta  \\
  \Lambda \cos\theta & -\frac{\Lambda}{\rho}\sin\theta & \Lambda X_0 \cos\theta  \\
  \Lambda \frac{\rho^2 +1}{\rho X_0} \cos\theta & -\Lambda \frac{\rho^2 +1}{\rho^2 X_0} \cos\theta & -\Lambda \frac{1-\rho^2}{\rho} \sin \theta
  \end{array}
  \right] \nonumber
\end{equation}
\begin{equation}
B_{3\times2}=\left[
  \begin{array}{cc}
  X_0 & 0 \\
  0 & \rho X_0 \\
  \alpha & -\alpha 
  \end{array}
  \right] \nonumber
\end{equation}
\begin{equation}
C_{2\times3}=\left[
  \begin{array}{ccc}
  -\frac{2}{T}(1+2 \Lambda \rho \sin\theta)X_0 & 0 & 0 \\
  0 & -\frac{2}{T}(1-2\frac{\Lambda}{\rho}\sin\theta) \rho X_0 & 0 
  \end{array}
  \right] \nonumber
\end{equation}
\begin{equation}
D_{2\times2}=\left[
  \begin{array}{cc}
  -\frac{1}{T}(1+2X_0^2) & 0 \\
   0 & -\frac{1}{T}(1+2\rho^2 X_0^2)
  \end{array}
  \right] \nonumber
\end{equation}
Stability requires that all eigenvalues have a nonpositive real part.  At the boundaries of the stability regions, the system undergoes Hopf bifurcations giving rise to stable limit cycles corresponding to undamped relaxation oscillations characterized by asymmetric synchronized oscillations of the electric fields, that can have different mean values and amplitudes \cite{Kovanis_97}. Exceptional Points correspond to points of the parameter space where two (or more) eigenvalues coalesce and can lie either in the stable or the unstable region of the solution and parameter space. \

\begin{figure}[h!]
  \begin{center}
  \hspace{-1em}{\scalebox{\scl}{\includegraphics{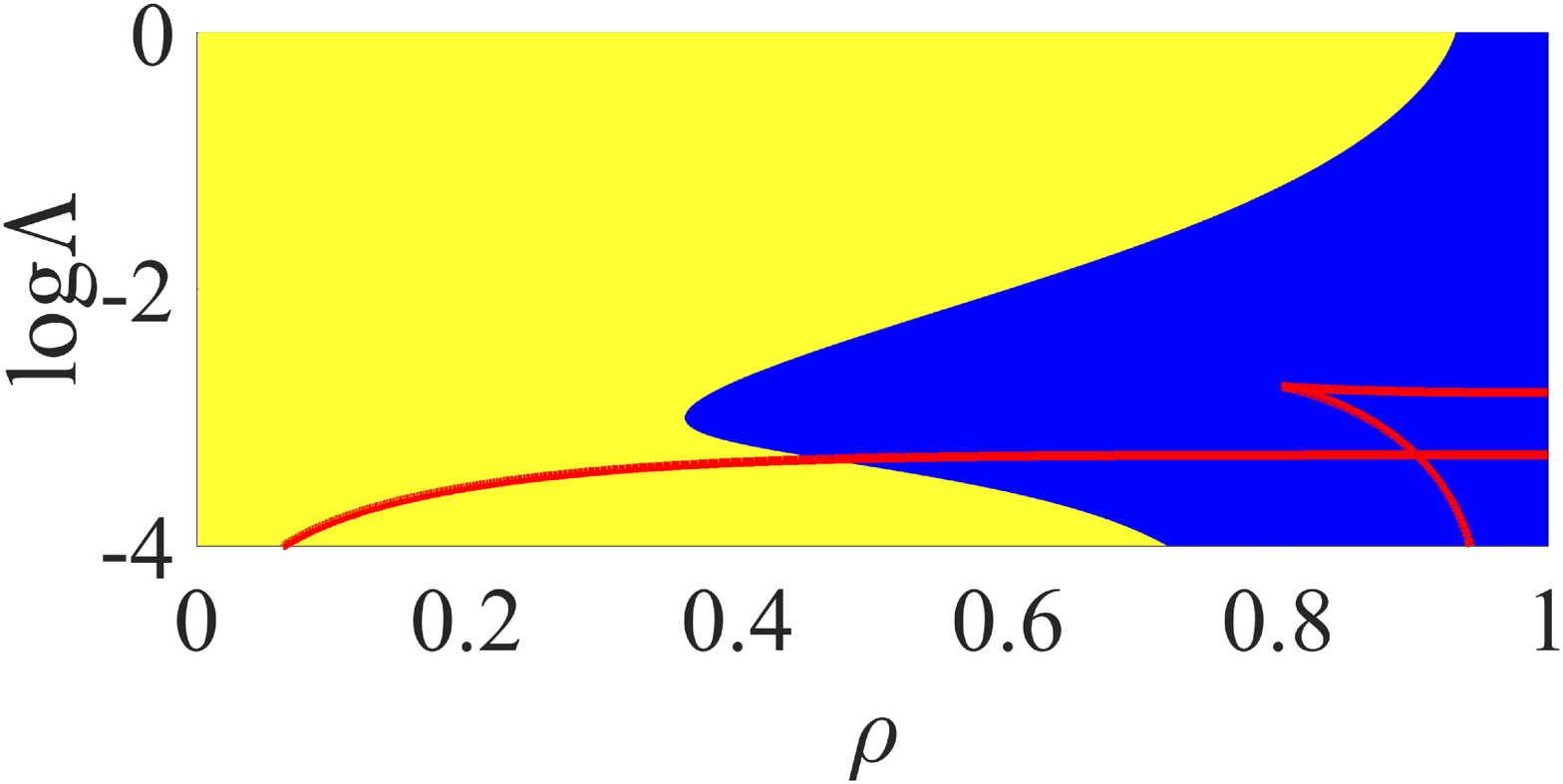}}}
  \hspace{-1em}{\scalebox{\scl}{\includegraphics{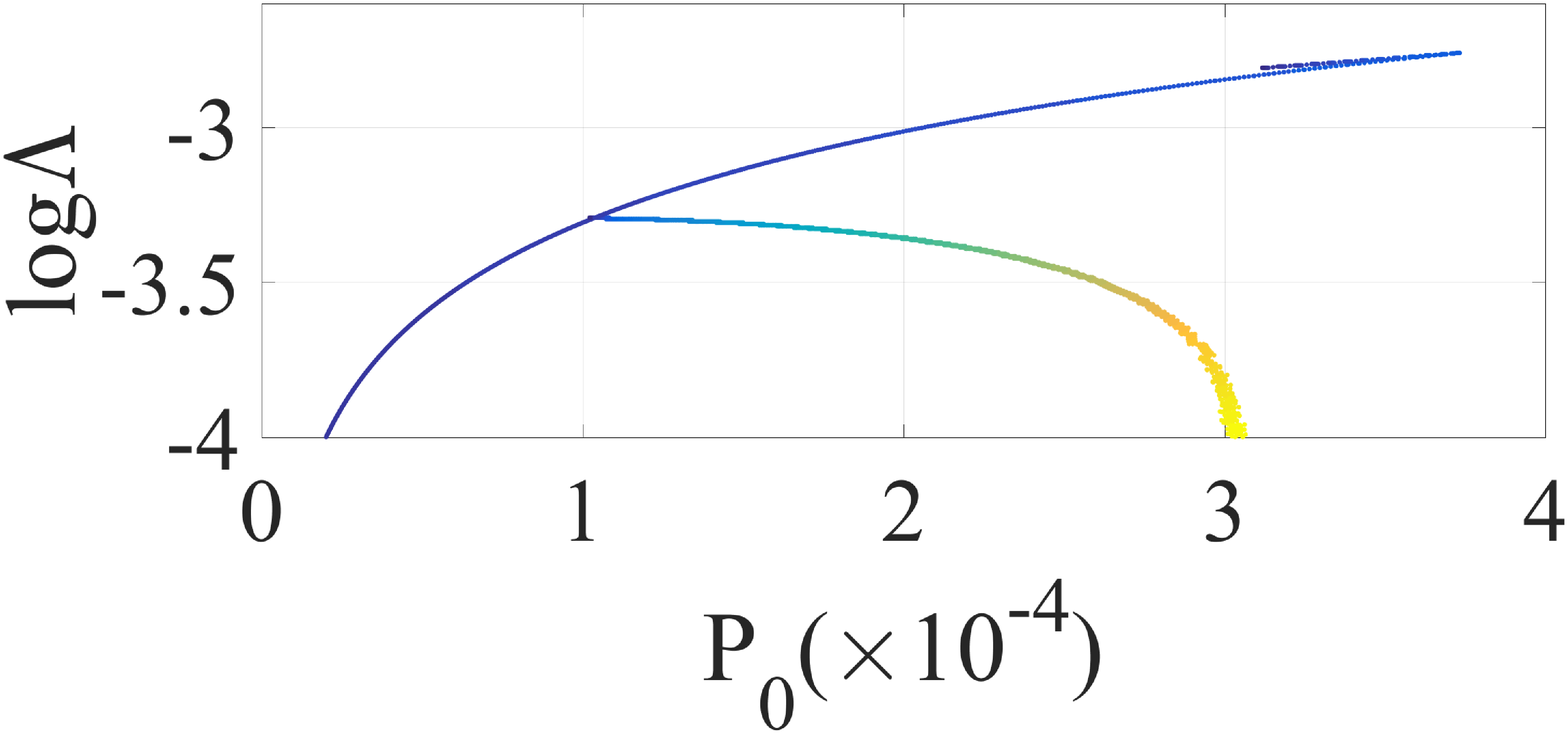}}}
  \caption{Stability and location of Hopf Bifurcation and Exceptional Points in the $(\Lambda, \rho)$ (left) and $(\Lambda, P_0)$ (right) space, for the case of zero frequency detuning ($\Delta=0$) and equal pumping ($P_1=P_2=P_0$). (left) The Hopf Bifurcation Points are located at the boundary between stable (blue) and unstable (yellow) regions. Red lines depict the location of the Exceptional Points. (right) The line colormap depicts the asymmetry of the respective phase-locked states (blue and yellow color correspond to $\rho=1$ and $\rho<1$, respectively). The reference electric field amplitude $(X_0)$ is given by (\ref{X0_rho}). }
  \end{center}
\end{figure}

In what follows we focus on a system with linewidth enhancement factor $\alpha=5$ and carrier to photon lifetime ratio $T=400$. For a zero detuning $(\Delta=0)$ and equal pumping $(P_1=P_2=P_0)$ there exists one asymmetric phase-locked state with arbitrary electric field amplitude ration $(\rho)$. The phase difference and reference field amplitude are given by Eq. (\ref{theta_rho}) and (\ref{X0_rho}), with the different $s=0,1$ corresponding to a trivial exchange of indices between the two lasers. The stability domain of these states in parameter-solution space $(\Lambda, \rho)$ is depicted in Fig. 1(left) with the Hopf bifurcation points corresponding to the boundary between the stable and the unstable region and the exceptional points depicted by a red line. It is worth emphasizing that exceptional points can be located either in the stability or in the instability region. The location of the exceptional points in the parameter space $(\Lambda, P_0)$ is depicted in Fig. 1(right) with the line colormap corresponding to different values of phase-locked state asymmetry (blue and yellow color correspond to $\rho=1$ and $\rho<1$, respectively). This case corresponds to the aforementioned ``missed'' asymmetric phase-locked state in a fully symmetric configuration examined in \cite{Winful&Wang_88} and shows clearly the relation between the existence of Exceptional Points and the inherently asymmetric dynamics of the system . \

\begin{figure}[h!]
  \begin{center}
  \hspace{-1em}{\scalebox{\scl}{\includegraphics{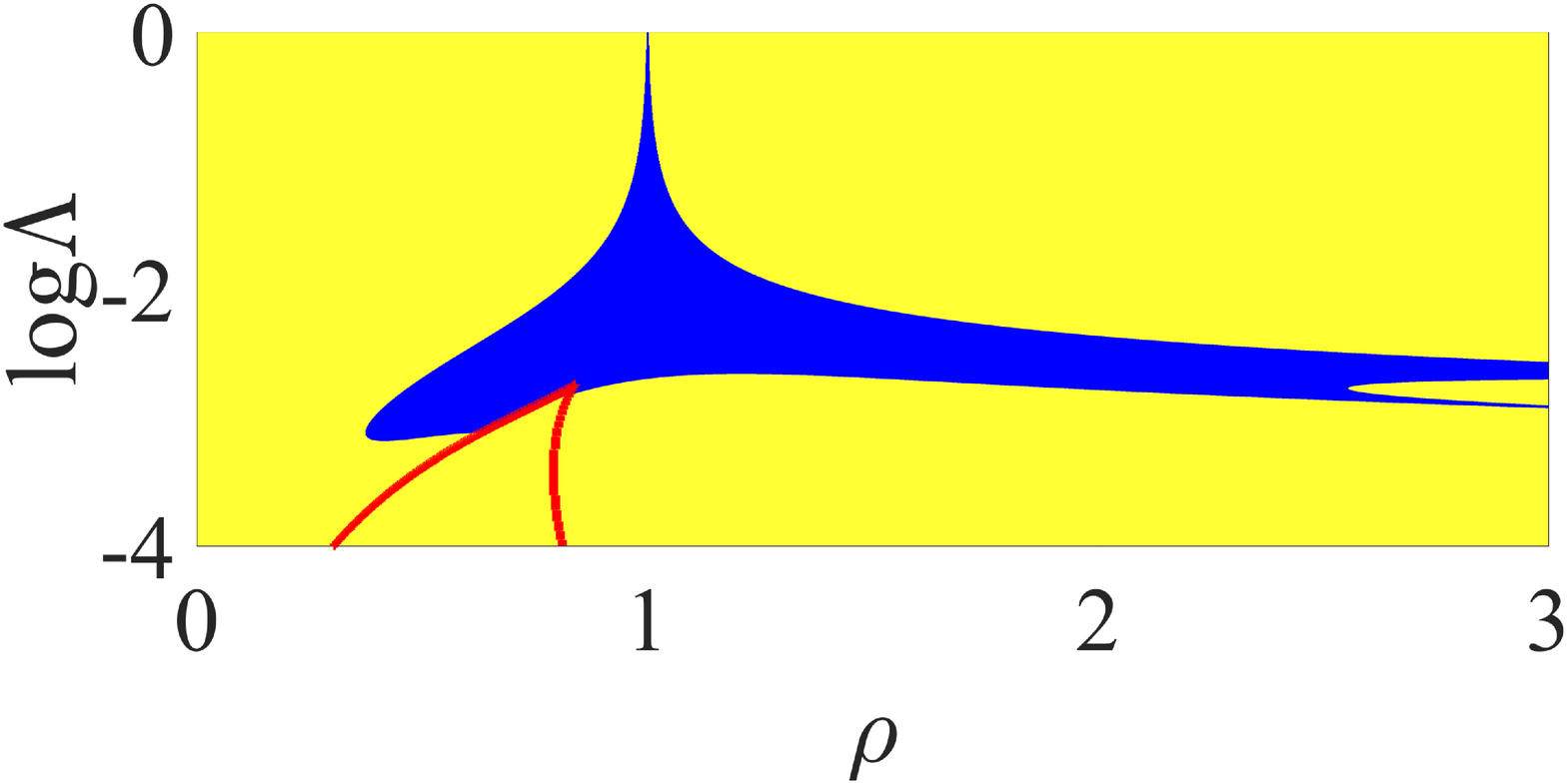}}}
  \hspace{-1em}{\scalebox{\scl}{\includegraphics{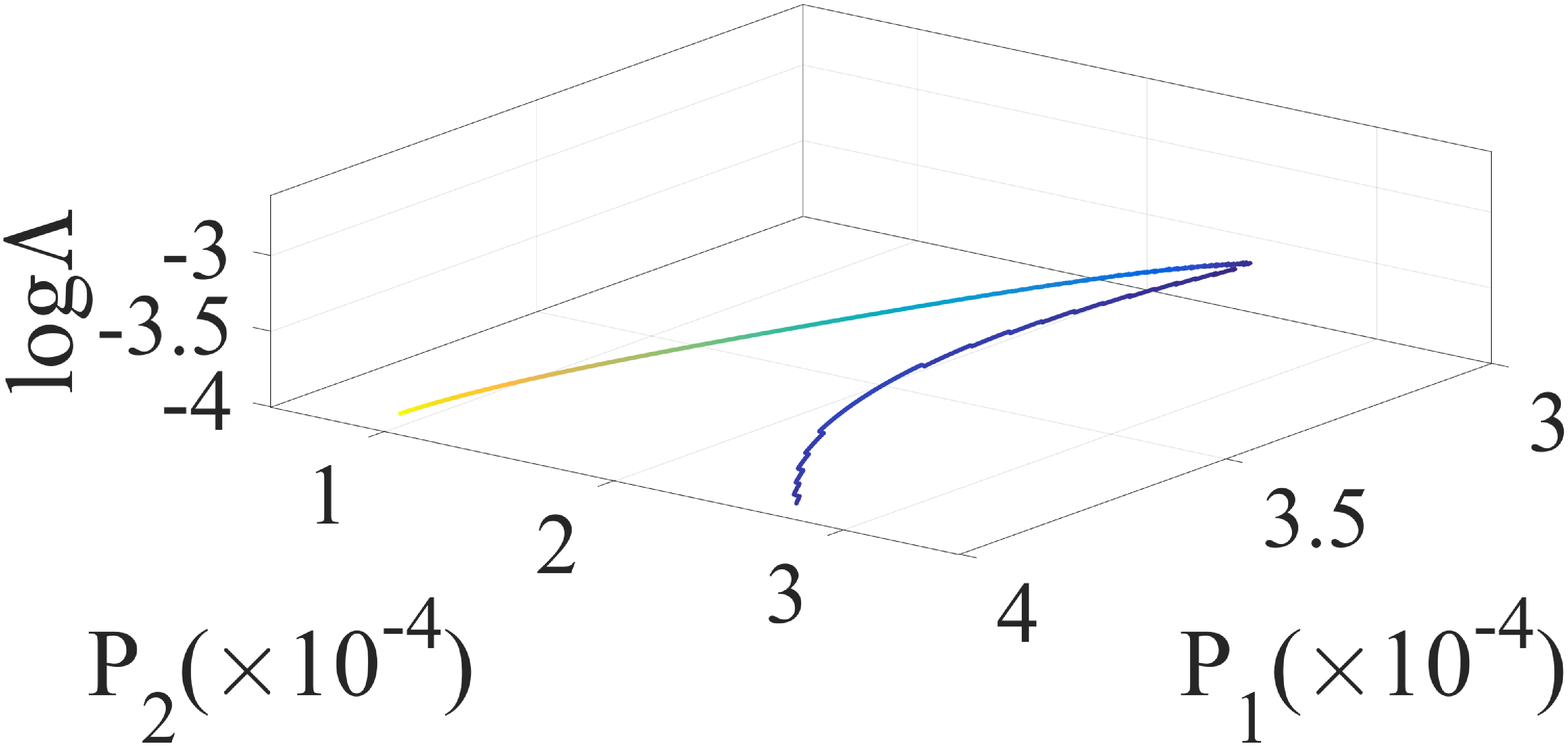}}}\\
  \hspace{-1em}{\scalebox{\scl}{\includegraphics{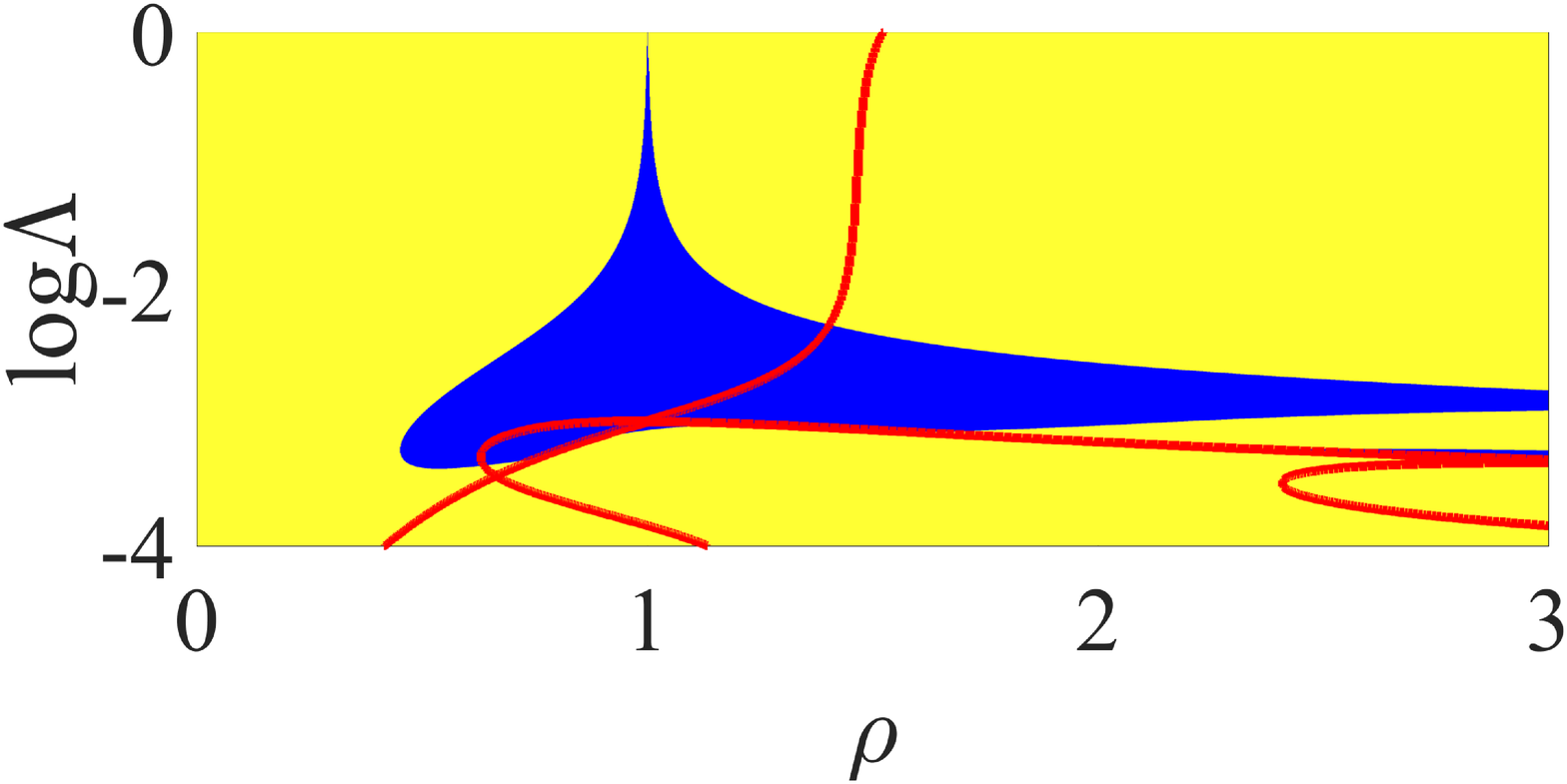}}}
  \hspace{-1em}{\scalebox{\scl}{\includegraphics{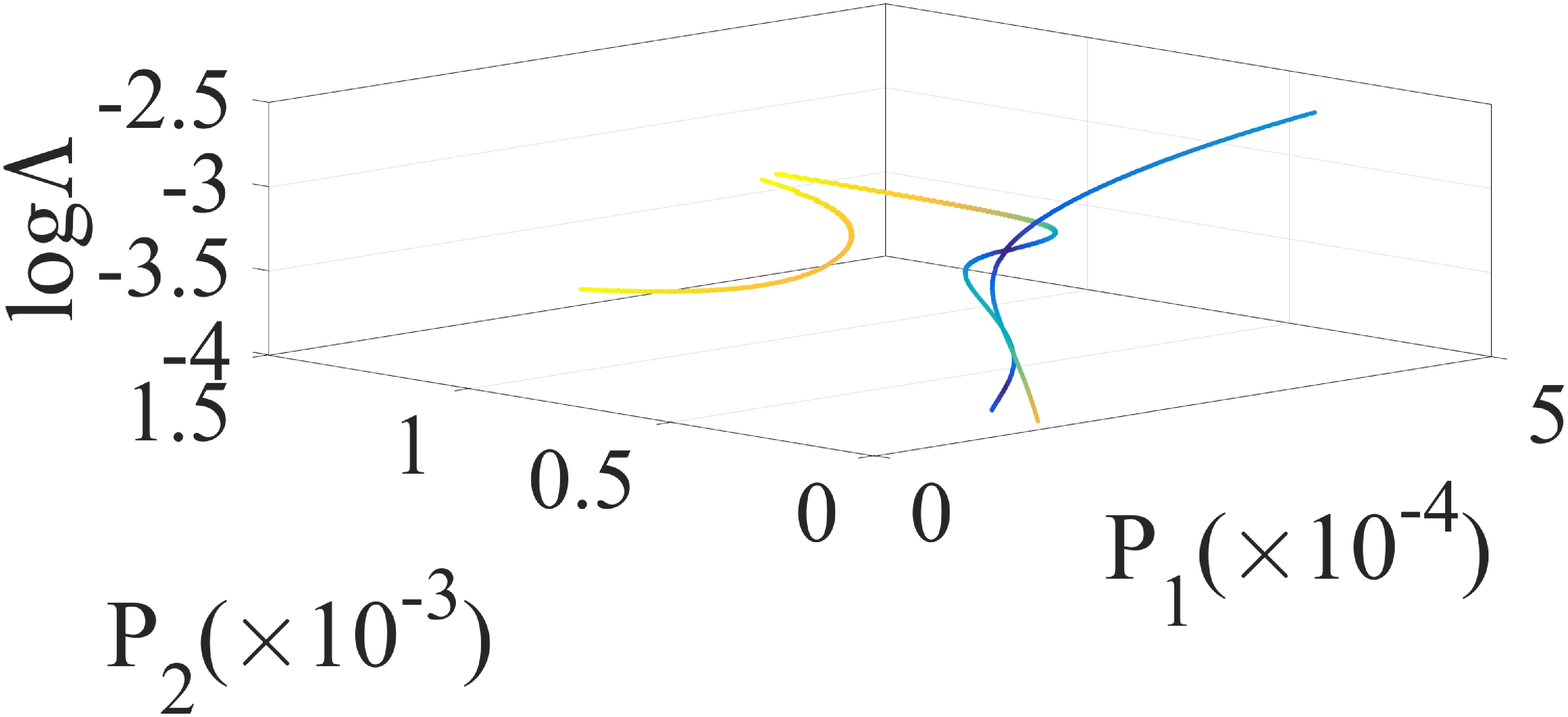}}}
  \caption{Stability and location of Hopf Bifurcation and Exceptional Points in the $(\Lambda, \rho)$ (left) and $(\Lambda, P_1, P_2)$ (right) space, for the case of zero frequency detuning ($\Delta=0$) and unequal pumping ($P_1\neq P_2$). (left) The Hopf Bifurcation Points are located at the boundary between stable (blue) and unstable (yellow) regions. Red lines depict the location of the Exceptional Points. (right) The line colormap depicts the asymmetry of the respective phase-locked states (blue and yellow color correspond to $\rho=1$ and $\rho<1$ or $\rho>1$, respectively). The phase difference $(\theta)$ is given by (\ref{theta_rho}) with $s=0$, and the reference electric field amplitude of the corresponding phase-locke modes is $X_0=10^{-1.7}$ (top) and $X_0=10^{-1.9}$ (bottom). The topology of the lines of Exceptional Points depends crucially on the total power related to $X_0$.}
  \end{center}
\end{figure}

\begin{figure}[h!]
  \begin{center}
  \hspace{-1em}{\scalebox{\scl}{\includegraphics{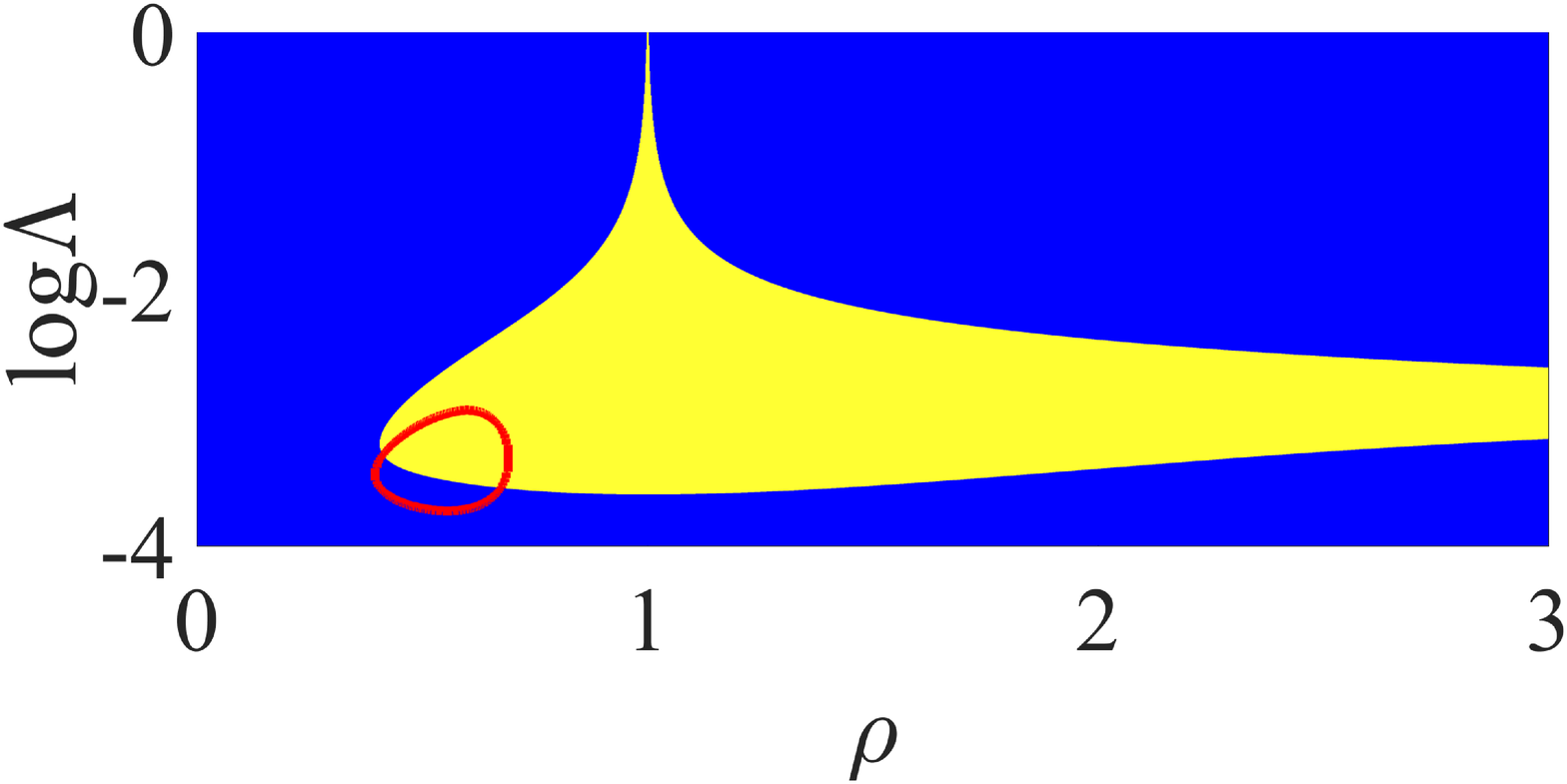}}}
  \hspace{-1em}{\scalebox{\scl}{\includegraphics{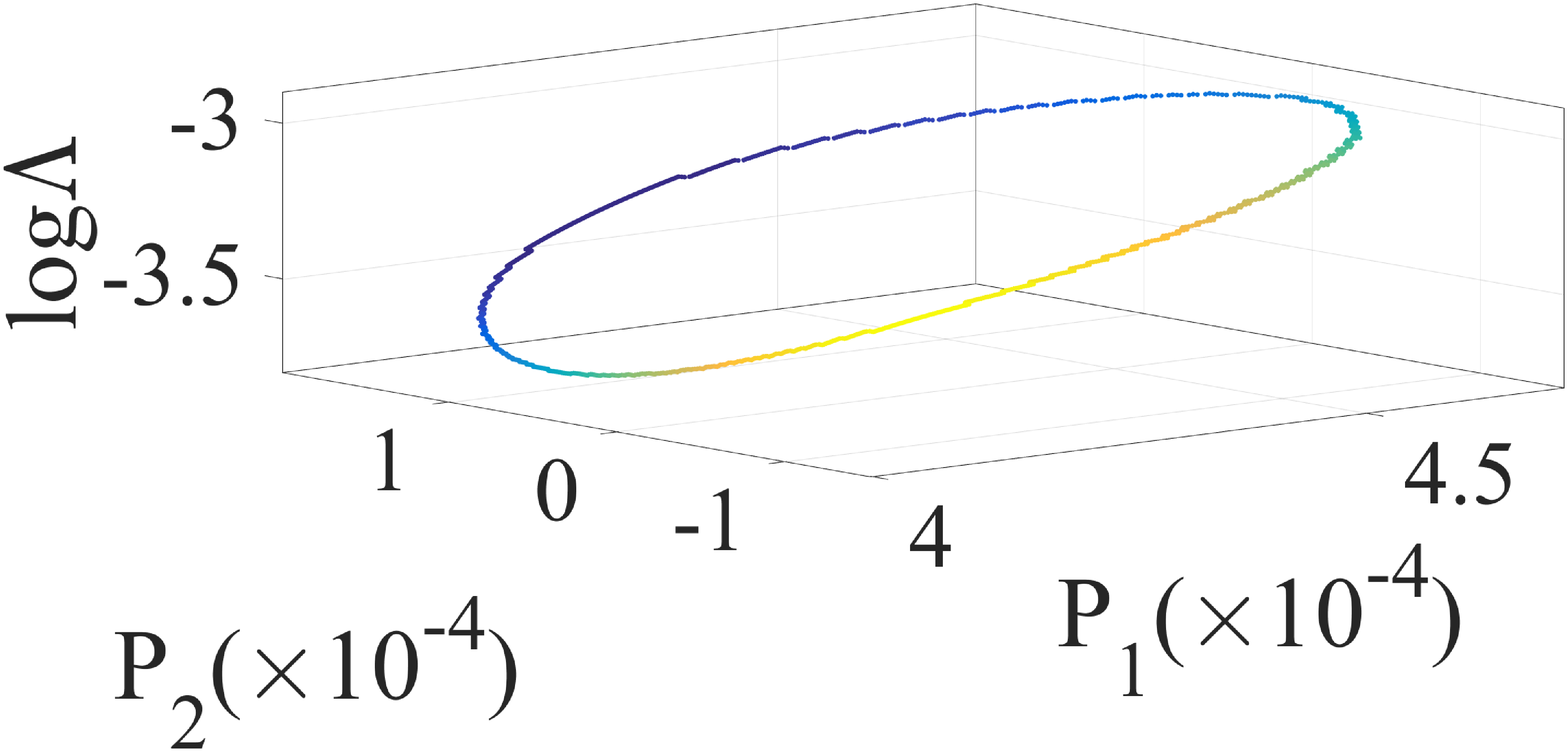}}}\\
  \hspace{-1em}{\scalebox{\scl}{\includegraphics{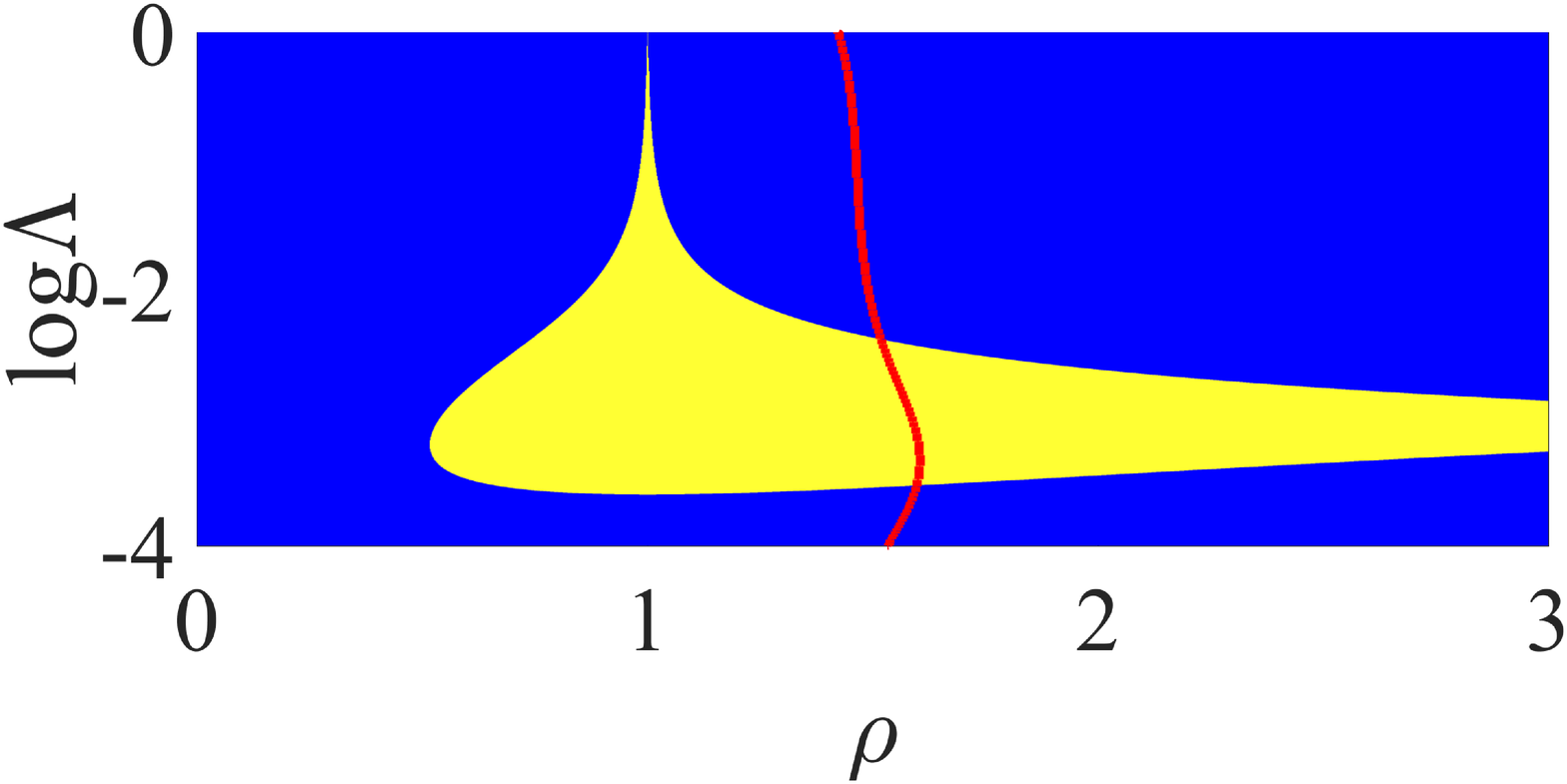}}}
  \hspace{-1em}{\scalebox{\scl}{\includegraphics{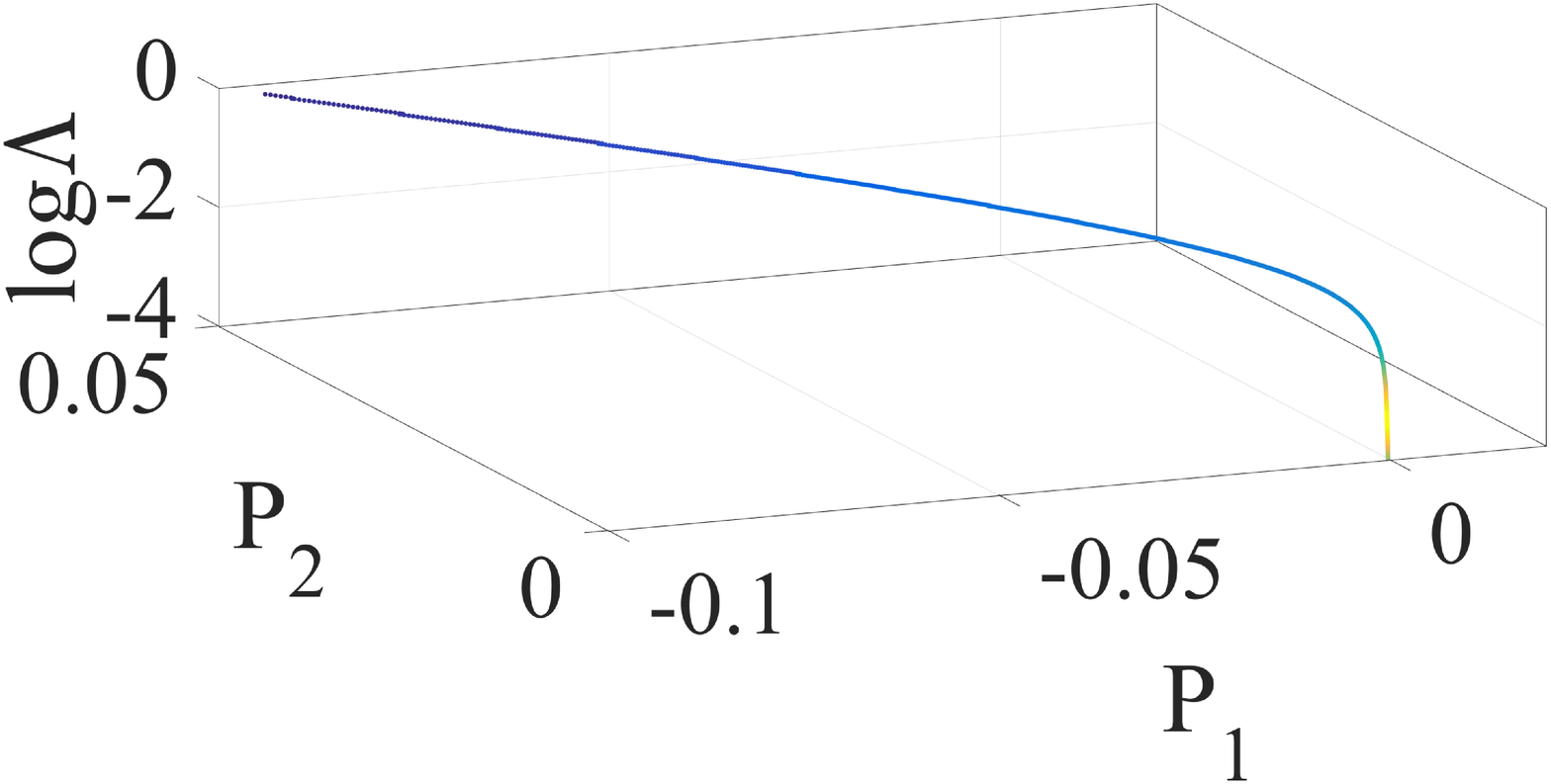}}}
  \caption{Stability and location of Hopf Bifurcation and Exceptional Points in the $(\Lambda, \rho)$ (left) and $(\Lambda, P_1, P_2)$ (right) space, for the case of zero frequency detuning ($\Delta=0$) and unequal pumping ($P_1\neq P_2$). (left) The Hopf Bifurcation Points are located at the boundary between stable (blue) and unstable (yellow) regions. Red lines depict the location of the Exceptional Points. (right) The line colormap depicts the asymmetry of the respective phase-locked states (blue and yellow color correspond to $\rho=1$ and $\rho<1$ or $\rho>1$, respectively). The phase difference $(\theta)$ is given by (\ref{theta_rho}) with $s=1$, and the reference electric field amplitude of the corresponding phase-locke modes is $X_0=10^{-1.7}$ (top) and $X_0=10^{-1.9}$ (bottom). The topology of the lines of Exceptional Points depends crucially on the total power related to $X_0$.}
  \end{center}
\end{figure}

For the case of zero detuning $(\Delta=0)$ and unequal pumping $(P_1 \neq P_2)$, for every value of $\rho$ there exist two asymmetric phase-locked states with phase differences given by Eq. (\ref{theta_rho}) for $s=0,1$, respectively. Their stability as well as the location of the Hopf bifurcation and exceptional points are depicted in Figs. 2, 3(left) for different values of $X_0$ which is not fixed in this case. Moreover, the parameter space has an additional dimension $(\Lambda, P_1, P_2)$ and the location of the exceptional points in this space is depicted in Figs. 2, 3(right). A qualitativily different topology of the lines of exceptional points, is shown, including open and closed lines as well as multiple intersecting lines, depending on the total electric field power related to $X_0$.\

\begin{figure}[pt]
  \begin{center}
  \hspace{-1em}{\scalebox{\scl}{\includegraphics{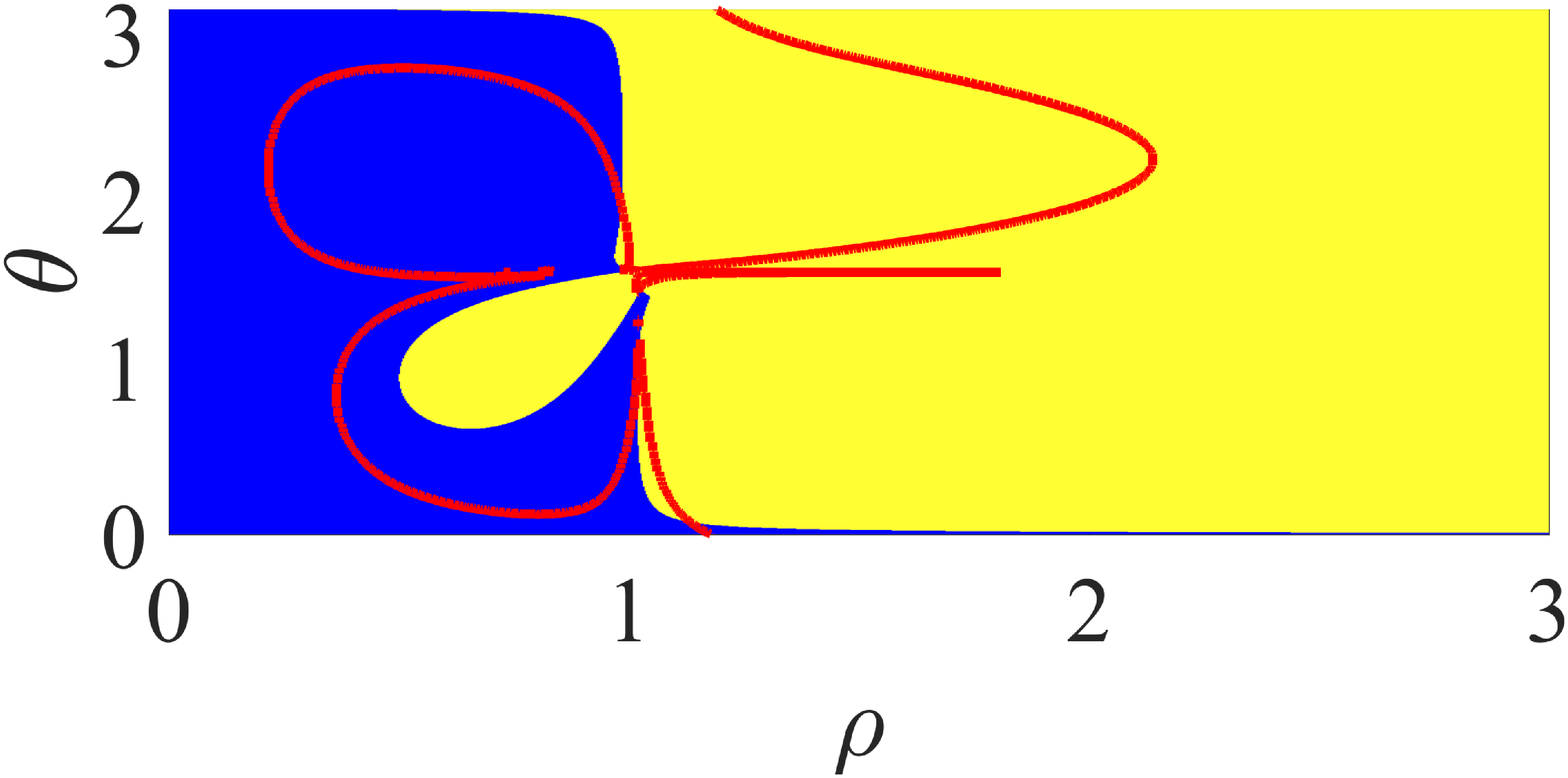}}}
  \hspace{-1em}{\scalebox{\scl}{\includegraphics{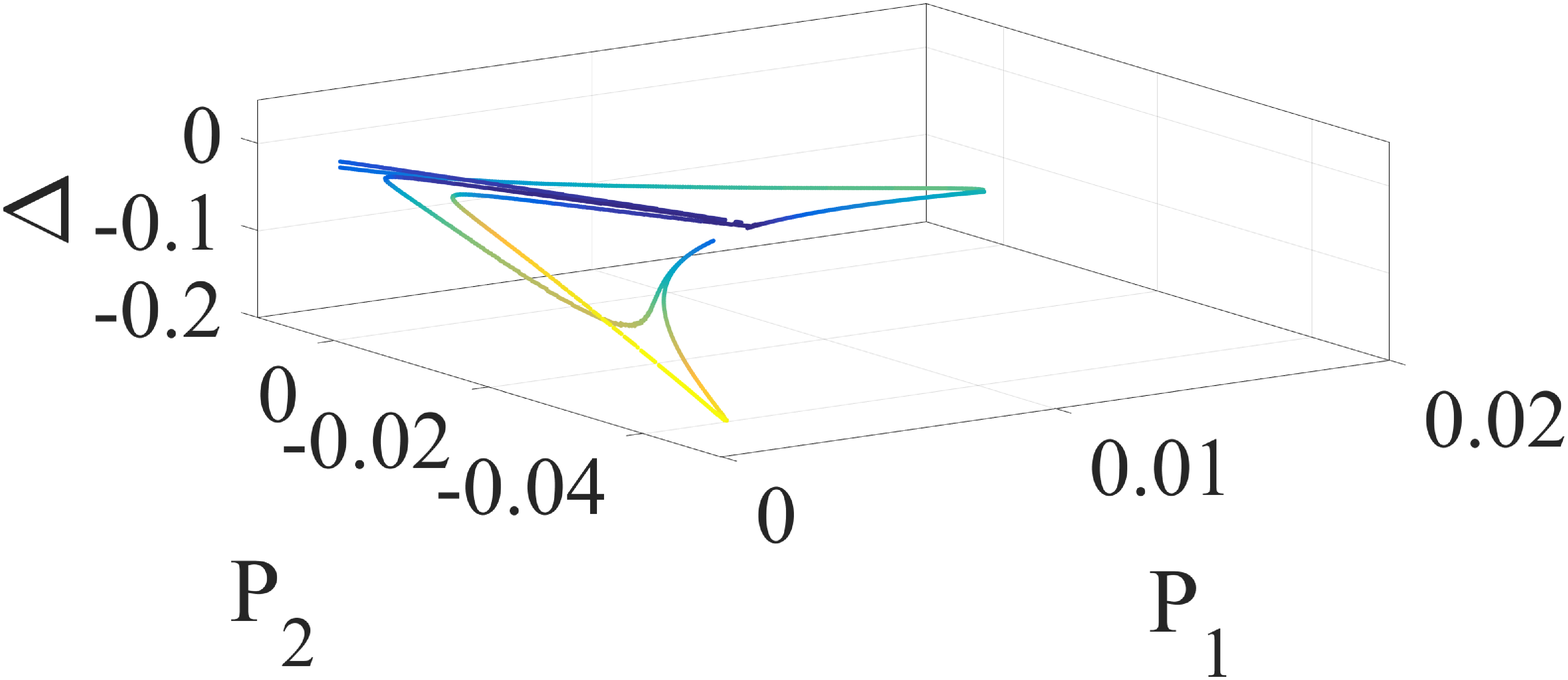}}}\\
  \hspace{-1em}{\scalebox{\scl}{\includegraphics{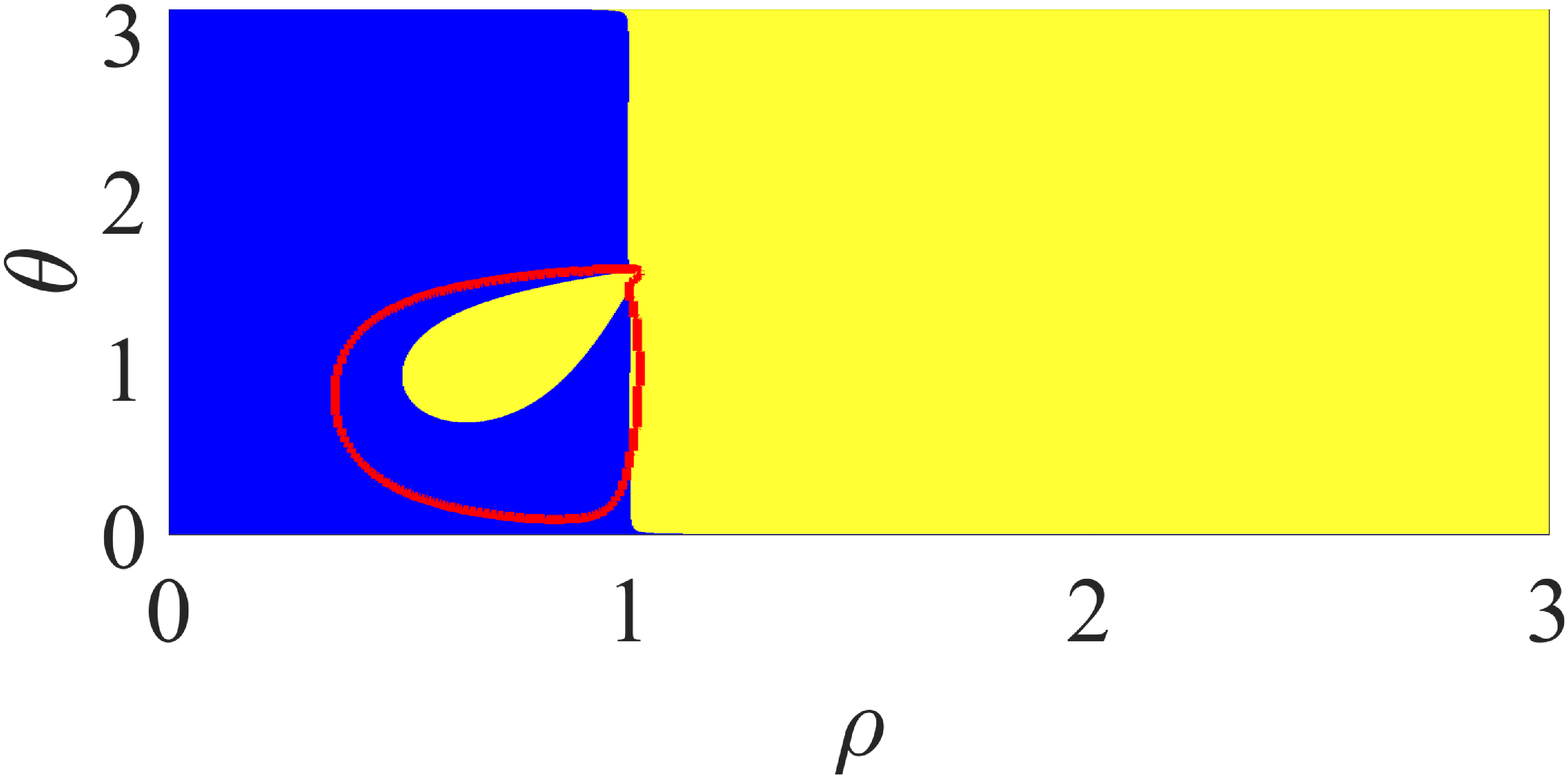}}}
  \hspace{-1em}{\scalebox{\scl}{\includegraphics{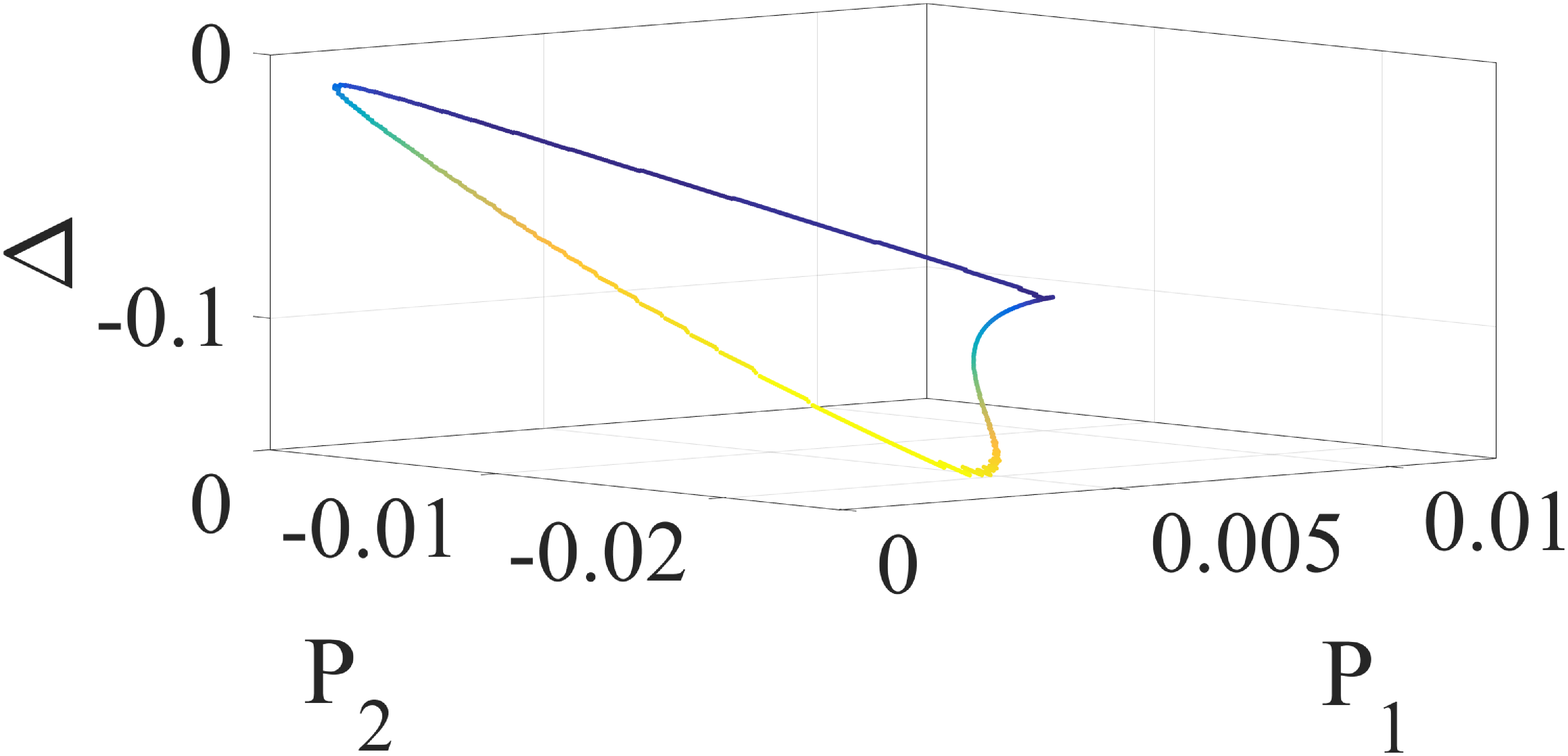}}}\\
  \hspace{-1em}{\scalebox{\scl}{\includegraphics{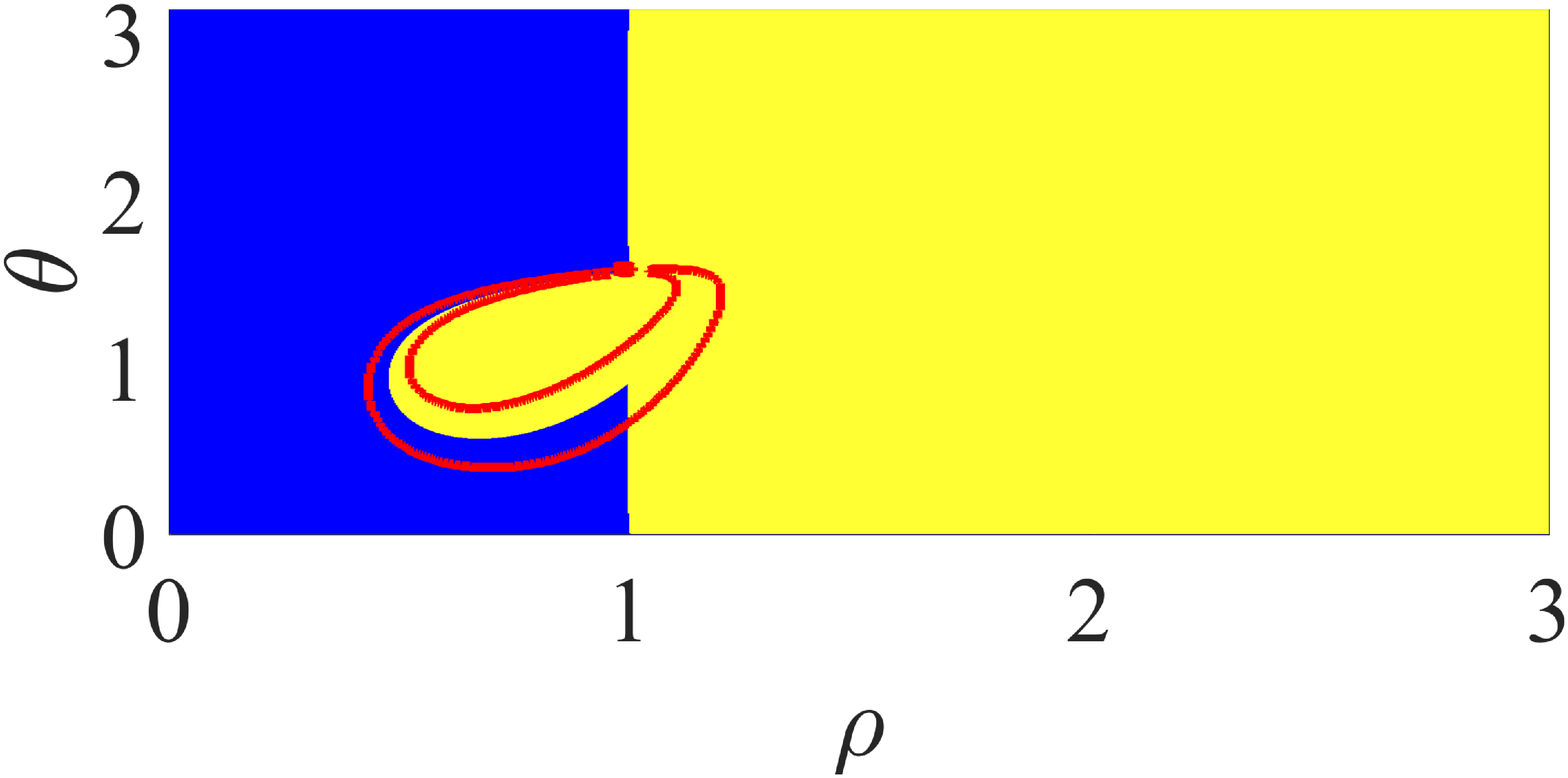}}}
  \hspace{-1em}{\scalebox{\scl}{\includegraphics{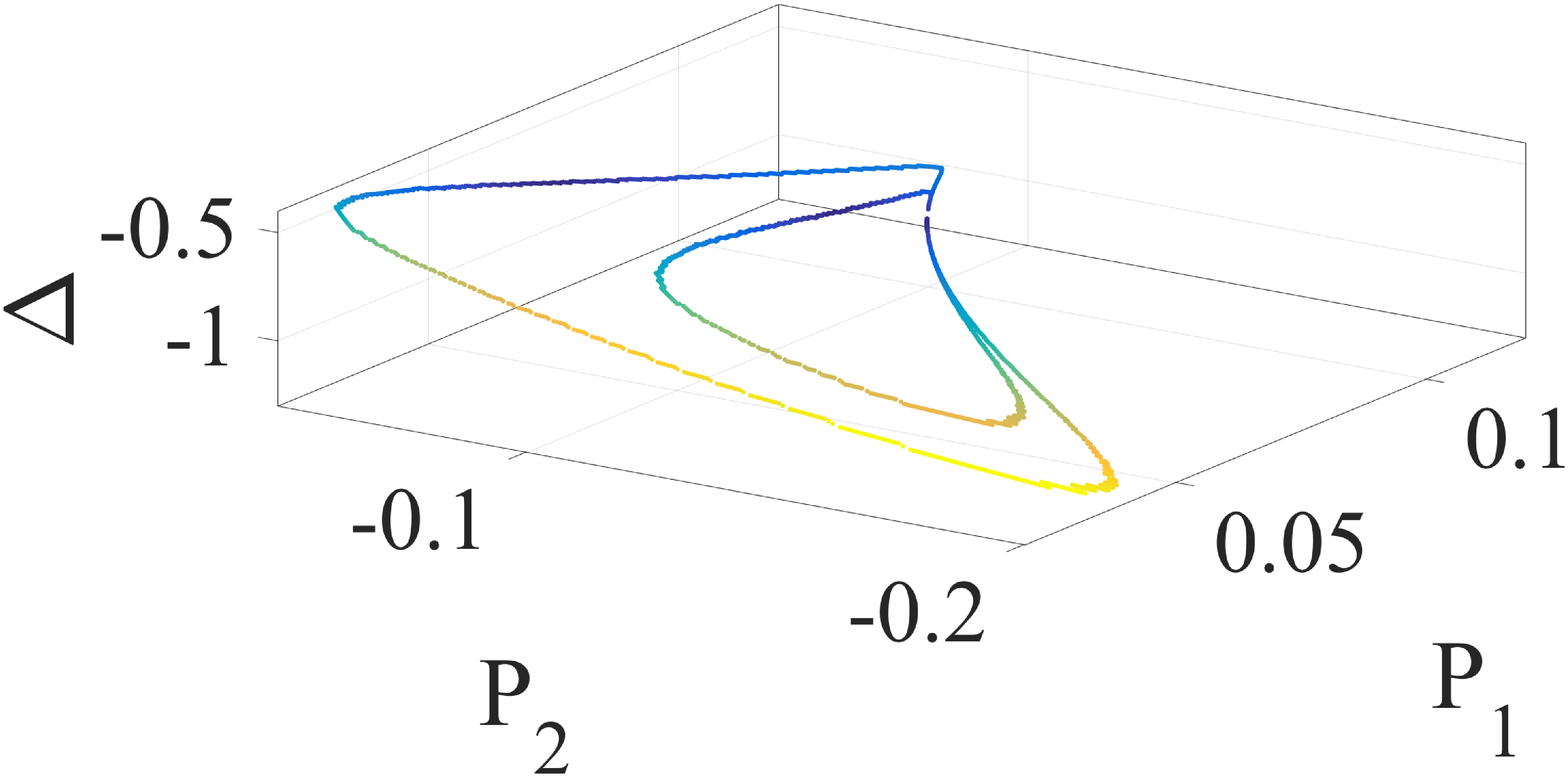}}}
  \caption{Stability and location of Hopf Bifurcation and Exceptional Points in the $(\theta, \rho)$ (left) and $(\Delta, P_1, P_2)$ (right) space, for the case of non-zero frequency detuning ($\Delta \neq 0$) and unequal pumping ($P_1\neq P_2$). (left) The Hopf Bifurcation Points are located at the boundary between stable (blue) and unstable (yellow) regions. Red lines depict the location of the Exceptional Points. (right) The line colormap depicts the asymmetry of the respective phase-locked states (blue and yellow color correspond to $\rho=1$ and $\rho<1$ or $\rho>1$, respectively). The normalized coupling $(\Lambda)$ and the reference electric field amplitude of the corresponding phase-locke modes are: $\Lambda=10^{-1}, X_0=10^{-1.8}$ (top), $\Lambda=10^{-2}, X_0=10^{-2.5}$ (middle), and $\Lambda=10^{-1}, X_0=10^{-2.5}$ (bottom).  }
  \end{center}
\end{figure}

The most general case corresponds to configurations where we have non-zero detuning $(\Delta \neq 0)$ and unequal pumping $(P_1 \neq P_2)$, shown in Fig. 4. The lines of exceptional points can by located either in the stability or in the instability regions and may encircle the Hopf bifurcation lines in the $(\Lambda, \rho)$ space [Fig. 4(left)]. The parameter space is four dimensional $(\Lambda, \Delta, P_1, P_2)$ and the location of the exceptional points is depicted in the subspace $(\Delta, P_1, P_2)$ for different values of $\Lambda$ in Fig. 4(right). It is worth mentioning that knoweledge of the location of the exceptional points in this parameter space is of particular importance since in most realistic configuration both the pumping and the optical detuning are controlled by the same experimental parameter, i.e. the injection current, so that the system parameters can be chosen along specific paths in the parameter space \cite{Choquette_18}. In such cases the relative position of these experimentally accessible paths with respect to the lines of exceptional points is crucial for the judicious configuration for operation at  an Exceptional Point or in its vicinity.
\vspace{1.5em}

\section{Conclunding remarks}
We have studied the properties of the eigenvalue spectrum of asymmetric phase-locked modes in coupled semiconductor lasers. The stability and spectral line shape of theses modes are determined by the Hopf bifurcation and the exceptional points, the locations of which are shown in both the solution and the parameter space. It has been shown that the existence of exceptional points is not restricted either to PT-symmetric or effectivelly PT-symmetric conditions. The complex topology of the lines of exceptional points in the parameter space suggests that there is enough flexibility for the judicious selection of parameters, such as frequency detuning and pumping, for operation exactly at, or at the vicinity of an exceptional point by taking into account issues of experimental constraints in the parameter space.

\section*{Acknowledgements}
This research is partly supported by two ORAU grants entitled "Taming Chimeras to Achieve the Superradiant Emitter" and ''Dissecting the Collective Dynamics of Arrays of Superconducting Circuits and Quantum Metamaterials'', funded by Nazarbayev University. Funding from MES RK state-targeted program BR05236454 is also acknowledged.

\end{document}